

\font\twelverm=cmr10 scaled 1200    \font\twelvei=cmmi10 scaled 1200
\font\twelvesy=cmsy10 scaled 1200   \font\twelveex=cmex10 scaled 1200
\font\twelvebf=cmbx10 scaled 1200   \font\twelvesl=cmsl10 scaled 1200
\font\twelvett=cmtt10 scaled 1200   \font\twelveit=cmti10 scaled 1200
\font\twelvesc=cmcsc10 scaled 1200
\skewchar\twelvei='177   \skewchar\twelvesy='60
\def\twelvepoint{\normalbaselineskip=12.4pt
  \abovedisplayskip 12.4pt plus 3pt minus 6pt
  \belowdisplayskip 12.4pt plus 3pt minus 6pt
  \medskipamount=7.2pt plus2.4pt minus2.4pt
  \def\rm{\fam0\twelverm}          \def\it{\fam\itfam\twelveit}%
  \def\sl{\fam\slfam\twelvesl}     \def\bf{\fam\bffam\twelvebf}%
  \def\mit{\fam 1}                 \def\cal{\fam 2}%
  \def\tt{\twelvett}
  \def\sc{\twelvesc}
  \def\nullspace{\nulldelimiterspace=0pt \mathsurround=0pt }
  \def\big##1{{\hbox{$\left##1\vbox to 10.2pt{}\right.\nullspace$}}}
  \def\Big##1{{\hbox{$\left##1\vbox to 13.8pt{}\right.\nullspace$}}}
  \def\bigg##1{{\hbox{$\left##1\vbox to 17.4pt{}\right.\nullspace$}}}
  \def\Bigg##1{{\hbox{$\left##1\vbox to 21.0pt{}\right.\nullspace$}}}
  \textfont0=\twelverm   \scriptfont0=\tenrm   \scriptscriptfont0=\sevenrm
  \textfont1=\twelvei    \scriptfont1=\teni    \scriptscriptfont1=\seveni
  \textfont2=\twelvesy   \scriptfont2=\tensy   \scriptscriptfont2=\sevensy
  \textfont3=\twelveex   \scriptfont3=\twelveex  \scriptscriptfont3=\twelveex
  \textfont\itfam=\twelveit
  \textfont\slfam=\twelvesl
  \textfont\bffam=\twelvebf \scriptfont\bffam=\tenbf
  \scriptscriptfont\bffam=\sevenbf
  \normalbaselines\rm}

\def\beginlinemode{\endmode
  \begingroup\parskip=0pt \obeylines\def\\{\par}\def\endmode{\par\endgroup}}
\def\beginparmode{\endmode \begingroup \def\endmode{\par\endgroup}}
\let\endmode=\par
{\obeylines\gdef\
{}}
\def\singlespace{\baselineskip=\normalbaselineskip}

\def\doublespace{\baselineskip=\normalbaselineskip \multiply\baselineskip by 2}
\def\triplespace{\baselineskip=\normalbaselineskip \multiply\baselineskip by 3}
\newcount\firstpageno \firstpageno=2
\footline={\ifnum\pageno<\firstpageno{\hfil}\else{\hfil
                                                  \twelverm\folio\hfil}\fi}
\let\rawfootnote=\footnote		
\def\footnote#1#2{{\rm\parindent=20pt\singlespace\hang
  \rawfootnote{#1}{\tenrm#2\hfill\vrule height 0pt depth 6pt width 0pt}}}
\def\raggedcenter{\leftskip=4em plus 12em \rightskip=\leftskip
  \parindent=0pt \parfillskip=0pt \spaceskip=.3333em \xspaceskip=.5em
  \pretolerance=9999 \tolerance=9999 \hyphenpenalty=9999 \exhyphenpenalty=9999
}
\parskip=\medskipamount \twelvepoint \overfullrule=0pt
\font \titlefont=cmr10 scaled \magstep4
\def\bigtitle{\null\vskip 3pt plus 0.3fill \beginlinemode
   \triplespace \raggedcenter \titlefont}
\def\endtitlepage{\vfill\eject\beginparmode}

\def\half{{\textstyle{ 1\over 2}}}
\def\frac#1#2{{\textstyle{#1\over #2}}}

\def\sss{\scriptscriptstyle}
\def\gtwid{\mathrel{\raise.3ex\hbox{$>$\kern-.75em\lower1ex\hbox{$\sim$}}}}
\def\ltwid{\mathrel{\raise.3ex\hbox{$<$\kern-.75em\lower1ex\hbox{$\sim$}}}}

\def\T{\widetilde{T}}
\def\S{S}
\def\s{\sigma}
\def\Sk{{\cal J}}
\def\n{\hat n}
\def\x{\zeta}
\def\dO{d\Omega}
\def\cd{\!\cdot\!}

\def\etal{et al.}
\def\la{\bigl\langle}
\def\ra{\bigr\rangle}
\def\Rp{R^{\,\prime}}
\def\vx{\la\x^2(\alpha)\ra}

\hyphenation{parameter}

\singlespace
\rightline{CfPA--93--th--18}
\rightline{astro-ph/9306012}
\rightline{June 1993}
\rightline{REVISED}

\doublespace

\bigtitle
Cosmic Variance of the Three-Point Correlation Function
of the Cosmic Microwave Background
\beginparmode

\vskip 3pt plus 0.3fill
\centerline{{\sc Mark Srednicki}\footnote{*}{E-mail:
marks@tpau.physics.ucsb.edu.  On leave from
Department of Physics, University of California, Santa Barbara, CA 93106.}}

\vskip 3pt plus 0.1fill
\centerline{\rm Center for Particle Astrophysics, University of California,
                Berkeley, CA 94720}

\vskip 3pt plus 0.3fill \rm
\centerline{ABSTRACT}
\baselineskip=15pt
\narrower
The fluctuations in the cosmic microwave background radiation may contain
deviations from gaussian statistics which would be reflected in a nonzero
value of three-point correlation function of $\Delta T$.  However, any
potential observation of the three-point function is limited by cosmic
variance, even if a whole-sky map of $\Delta T$ is available.  Here I derive
a general formula for the cosmic variance of the three-point function in
terms of integrals over the two-point function.  This formula can be applied
to any cosmological model and to any experimental measurement strategy.
It gives a fundamental lower limit on the magnitude of a measurable
three-point correlation function, and hence on the measurable
amount of skewness in the distribution of $\Delta T$.

\noindent
{\it Subject headings:} cosmic microwave background --- cosmology: theory

\endtitlepage
\baselineskip=15pt
Now that anisotropies in the cosmic microwave background radiation have
been discovered on large angular scales (Smoot \etal\ 1992), attention
can be focused on the detailed properties of these fluctuations.
A question of particular importance is whether or not their distribution
is gaussian.  A gaussian distribution has a vanishing three-point correlation
function.  An apparently simple thing to check is whether or not this is true
of the observed anisotropies (to within the experimental error bars).

However, theoretical predictions of properties of the CMBR are always
probabilistic in nature, and fundamentally limited by the ``cosmic variance''
which arises from our inability to make measurements in more than one universe
(e.g., Abbott \& Wise 1984b; Scaramella \& Vittorio 1990, 1991;
Cayon \etal\ 1991; White \etal\ 1993).  The prediction of a vanishing
three-point function is no exception, and so comes with a purely theoretical
error bar.

In this letter I address this problem.  Assuming that the fluctuations are
in fact gaussian, I compute the expected deviation from zero of the three-point
correlation function (which includes, as a special case, the skewness), due
solely to the effect of making measurements in a single universe.  The answer
is expressed in terms of the theoretical
two-point correlation function, which depends on
both the window function of the experiment and the cosmological model which
is adopted.  I give detailed results for the COBE window function, assuming a
scale invariant ($n=1$) power spectrum for the fluctuations.  Any measured
three-point function must have a magnitude exceeding the values given here in
order to be construed as evidence for non-gaussian fluctuations.  I compare
my results for the cosmic variance of the skewness with previous Monte Carlo
simulations (Scaramella \& Vittorio 1991).

To begin, let $\T(\n)$ denote the temperature difference which the experiment
assigns to a point on the sky specified by the unit direction vector $\n$,
including the effects due to finite beam width and any ``chopping''
strategy which the experiment uses.  We can make a multipole expansion
of the form
$$\T(\n)=\sum_{lm}a_{lm}\,W_{lm}\,Y_{lm}(\n)\;,  \eqno(1)$$
where the $Y_{lm}(\n)$ are spherical harmonics,
the $W_{lm}$ represent the window function of the experiment, and
the $a_{lm}$ are random variables whose distribution must
be specified by a specific cosmological model.
In general, rotation invariance implies that
$$\la a_{lm} a^*_{l'm'}\ra = C_l\,\delta_{ll'}\,\delta_{mm'}\;,
                                                \eqno(2)$$
where the angle brackets denote an ensemble average over the
probability distribution for the $a_{lm}$.
Assuming an $n=1$ power spectrum and ignoring the possible
contribution of tensor modes, the Sachs--Wolfe effect results in
(Peebles 1982; Abbott \& Wise 1984a; Bond \& Efstathiou 1987)
$${C_l\over4\pi}={\frac65 Q^2\over l(l+1)}        \eqno(3)$$
for $l\ge2$, where $Q$ is the r.m.s.~ensemble average of the quadrupole moment
(which may differ significantly from the actual, measured quadrupole moment).
Since the monopole term must be removed from $\T(\n)$, we have $W_{00}=0$ and
$$\int {\dO\over 4\pi}\;\T(\n) = 0\;,  \eqno(4)$$
where $\dO$ denotes integration over the unit vector $\n$.
The theoretical two-point correlation function is given by
$$\eqalignno{
C_2(\n_1,\n_2) &= \la \T(\n_1)\T(\n_2) \ra \cr
\noalign{\medskip}
  &= \sum_{lm}C_l\,|W_{lm}|^2\,Y_{lm}(\n_1)\,Y^*_{lm}(\n_2)\;. &(5)\cr}$$
If $W_{lm}$ is independent of $m$, we can simplify this to
$$C_2(\n_1,\n_2) = C(x) \equiv {1\over4\pi}
                   \sum_{l}(2l+1)\,C_l\,|W_l|^2\,P_l(x)\;,\eqno(6)$$
where $x=\n_1\cd\n_2$ is the cosine of the beam-separation angle
and $P_l(x)$ is a Legendre polynomial.

Given the observed values of $\T(\n)$ for each $\n$, the experimental value
of the skewness $\S$ is given by the sky average of $\T^3$:
$$\S = \int{\dO\over4\pi}\;\T^3(\n)\;.\eqno(7)$$
(The effect of partial sky coverage will be discussed later.)
The corresponding dimensionless skewness parameter is
$\Sk = \S/\bigl[\int\dO\,\T^2(\n)/4\pi\bigr]{}^{3/2}$,
but a discussion of this quantity is complicated by the need to treat both
its numerator and denominator as random variables.  I therefore concentrate
on $S$ as defined in equation~(7).

If the distribution of the $a_{lm}$ is gaussian, then the ensemble average of
the skewness is zero: $\la\S\ra=0$.  Of course, for our particular sky the
experimental value of $\S$ is unlikely to be exactly zero, even if we
completely neglect experimental noise (Scaramella \& Vittorio 1991).  In order
to tell whether or not a particular measured value of $\S$ is significant
evidence of a departure from gaussian statistics, we need to know the cosmic
variance of $\S$, assuming that the underlying statistics are indeed gaussian.
Since the mean of $\S$ vanishes, we can write the variance of $\S$ as
$$\la\S^2\ra = \int{\dO_1\over4\pi}\,{\dO_2\over4\pi}\,
               \la\T^3(\n_1)\T^3(\n_2)\ra\;.\eqno(8)$$
Using standard combinatoric properties of gaussian distributions, we have
$$\la\T^3(\n_1)\T^3(\n_2)\ra
   = 9\,\la\T^2(\n_1)\ra \la\T^2(\n_2)\ra
        \la\T(\n_1)\T(\n_2)\ra
   + 6\,\la\T(\n_1)\T(\n_2)\ra^3\;.\eqno(9)$$
If $W_{lm}$ is independent of $m$, then ensemble averages are always
rotationally invariant, and $\la\T^2(\n_1)\ra$ is independent of
$\n_1$.  From now on, for simplicity, we will assume that this is the case.
Equation~(4) then implies that the first term in equation~(9) will vanish
after integration over $\n_1$.  Thus we have
$$\la\S^2\ra = 6 \int{\dO_1\over4\pi}\,{\dO_2\over4\pi}\,
                           C^3_2(\n_1,\n_2)\;.\eqno(10)$$
Making use of equation~(6), we can simplify this to
$$\la\S^2\ra = 3 \int_{-1}^{+1}dx\,C^3(x)\;.\eqno(11)$$

This formula assumes full sky coverage.
In the case of partial sky coverage, the analysis of Scott
\etal\ (1993) can be applied.  Specifically, the integrals in equation~(10)
should range only over the solid angle~$A$ which is covered, and each factor
of $4\pi$ should be replaced by~$A$.  Since $C_2(\n_1,\n_2)$ is always
sharply peaked near $\n_1=\n_2$, the net change is an enhancement of
$\la\S^2\ra$ by a factor of $4\pi/A$, provided that $A$ is large enough
to encompass the entire peak, and provided that equation~(4) still holds,
at least approximately.  For more details in the context of a different but
similar calculation, see Scott \etal\ (1993).

As a specific example, the COBE group (Smoot \etal\ 1992)
has reported results for the two-point correlation function
with the monopole, dipole, and quadrupole terms removed,
and the higher moments weighted with a $7^\circ\llap.5$~FWHM beam,
which results in $W_{lm}=W_l=\exp\bigl[-\half(l+\half)^2\s^2\bigr]$
with $\s=3^\circ\llap.2$ for $l>2$, and $W_l=0$ for $l\le2$.
Assuming equation~(3) for the moments of the temperature distribution
gives $C(1)=3.65\,Q^2$,
and performing the integral in equation~(11) numerically yields
$\la\S^2\ra = 1.10\,Q^6$ for full sky coverage.  Including the $l=2$ term
gives instead $C(1)=4.63\,Q^2$ and $\la\S^2\ra=3.44\,Q^6$.
In either case, a galactic latitude cut of $|b|>20^\circ$
enhances $\la\S^2\ra$ by a factor of $4\pi/A =1/(1-\cos70^\circ)=1.52$.

Thus, the skewness of the temperature distribution, as determined from the
COBE data with $|b|>20^\circ$ and the quadrupole removed,
would have to be significantly larger
than $\la\S^2\ra{}^{1/2}=1.3\,Q^3$ in order to constitute evidence
for non-gaussian fluctuations.  Here, again, $Q$ is the r.m.s.~ensemble
average of the quadrupole moment, which must be determined from the
normalization of the full two-point correlation function.  $Q$ is a convenient
measure of the overall magnitude of the fluctuations; for example, in
inflationary models, the amplitude of the perturbations at the time of
horizon crossing is given by $\varepsilon_{\rm\sss H} = (12/5\pi)^{1/2}Q/T_0$
(Abbott \& Wise 1984a,b).  If we assume an $n=1$ spectrum, then the COBE data
indicates that $Q=16.7 \pm 4\,\mu$K (Smoot \etal\ 1992).

Although the the probability distribution for $S$ must be symmetric about
$S=0$, it will not be gaussian.  (It is easy to check, for example, that
$\la\S^4\ra$ does not equal $3\la\S^2\ra{}^2$, as it would for a gaussian
distribution.) Thus, $\pm\la\S^2\ra{}^{1/2}$ does not correspond to a true
68\% confidence interval.  The probability distribution for the dimensionless
skewness parameter~$\Sk$ was computed via Monte Carlo methods by Scaramella
\& Vittorio (1991).  For an $n=1$ spectrum, beam width
$\sigma=3^\circ\llap.0$, and including the $l=2$ term, they found the $\Sk$
distribution to be well approximated by a gaussian with
$\la\Sk^2\ra{}^{1/2}=0.2$.  For comparison, we can compute
$\la\S^2\ra{}^{1/2}/C(1)^{3/2}$.  This is a ratio of averages rather than
the average of a ratio, so exact agreement is not expected.  Nevertheless,
in this case I find $C(1)=4.78\,Q^2$ and $\la\S^2\ra=3.56\,Q^6$, so that
$\la\S^2\ra{}^{1/2}/C(1)^{3/2}=0.18$, in good agreement with
Scaramella \& Vittorio's value of $\la\Sk^2\ra{}^{1/2}$.

Experiments on small angular scales typically result in a theoretical
two-point correlation function which can be adequately approximated by
$C(x)=C_0\exp\bigl[(x-1)/\theta_{\rm c}^2\bigr]$, where $C_0$ and
$\theta_{\rm c}$ are computable functions of the cosmological model and the
experimental parameters (e.g., Bond \& Efstathiou 1987).  In this case,
combining equation~(11) with the correction factor of Scott \etal\ (1993)
for partial sky coverage yields
$$\la\S^2\ra=(4\pi/A)\theta_{\rm c}^2 C_0^3\;.     \eqno(12)$$
Equation~(12) applies when $A\gg\theta_{\rm c}^2$; the upper limit on
$\la\S^2\ra$ is $6C_0^3$.

These results for the cosmic variance of the skewness can be extended
to the full three-point correlation function.
Experimentally, the three-point correlation function is determined
by first choosing a fixed configuration of three direction vectors;
the simplest choice is an equilateral triangle,
$$\n_1\cd\n_2 = \n_2\cd\n_3 = \n_3\cd\n_1 = \cos\alpha\;,  \eqno(13)$$
and we specialize to this case from here on.
The experimental three-point correlation function is then
$$\x(\alpha) = \int dR\;\;\T(R\n_1)\T(R\n_2)\T(R\n_3)\;.  \eqno(14)$$
Here the $\n_i$ are to be held fixed in a configuration obeying equation~(13),
and $R$ is a rotation matrix; appropriately integrating over $R$ results in
an average over all possible triangles obeying equation~(13).  Note that
$\x(0)$ is simply the skewness $\S$.  A specific realization of the $\n_i$ is
$$\eqalign{
\n_1 &= \bigl(s_\alpha, \quad 0, \quad c_\alpha \bigr)\;,\cr
\n_2 &= \bigl(-\half s_\alpha, \quad +\frac{\sqrt3}2s_\alpha,
                                \quad c_\alpha \bigr) \;,\cr
\n_3 &= \bigl(-\half s_\alpha, \quad -\frac{\sqrt3}2s_\alpha,
                                \quad c_\alpha \bigr) \;,\cr } \eqno(15) $$
where $c_\alpha=\bigl[(1+2\cos\alpha)/3\bigr]^{1/2}$ and
$s_\alpha=\bigl[(2-2\cos\alpha)/3\bigr]^{1/2}$.
This is an equilateral triangle centered on the $\hat z$ axis.
A specific realization of $R$ is
$$R = \varepsilon
      \pmatrix{ \cos\phi & \sin\phi & 0 \cr
\noalign{\medskip}
               -\sin\phi & \cos\phi & 0 \cr
\noalign{\medskip}
                       0 &        0 & 1 \cr }
      \pmatrix{ 1 &           0 &          0 \cr
\noalign{\medskip}
                0 &  \cos\theta & \sin\theta \cr
\noalign{\medskip}
                0 & -\sin\theta & \cos\theta \cr }
      \pmatrix{ \cos\psi & \sin\psi & 0 \cr
\noalign{\medskip}
               -\sin\psi & \cos\psi & 0 \cr
\noalign{\medskip}
                       0 &        0 & 1 \cr }.  \eqno(16)
$$
This is a product of a sign ($\varepsilon=\pm1$) and three rotations, the
first about the $\hat z$ axis by $\psi$, the second about the $\hat x$ axis
by $\theta$, and the third about the $\hat z$ axis by $\phi$.  The two
possible values of $\varepsilon$ must be summed over in order to get
triangles of both orientations.  The integration measure in equation~(14)
for this realization of $R$ is
$$\int dR = {1\over16\pi^2}\sum_{\varepsilon\,=\,\pm1}
            \int_0^\pi d\theta\,\sin\theta
            \int_0^{2\pi} d\phi
            \int_0^{2\pi} d\psi \;.  \eqno(17)$$

Assuming a gaussian distribution, the ensemble average of the three-point
correlation function vanishes: $\la\x(\alpha)\ra=0$.
The cosmic variance of $\x(\alpha)$ is then
$$\la\x^2(\alpha)\ra = \int dR\;\,d\Rp\; \la \T(R\n_1)\T(R\n_2)\T(R\n_3)
                           \T(\Rp\n_1)\T(\Rp\n_2)\T(\Rp\n_3)\ra\;.\eqno(18)$$
Using the rotation invariance of the ensemble average, the integral over
$\Rp$ can be removed, leaving
$$\la\x^2(\alpha)\ra = \int dR\;\, \la \T(R\n_1)\T(R\n_2)\T(R\n_3)
                           \T(\n_1)\T(\n_2)\T(\n_3) \ra \;.\eqno(19)$$
Using the appropriate generalization of equation~(9), symmetries of the
equilateral triangle, and $\int dR\;\,\T(R\n)=0$ [which follows from
equation~(4)], we ultimately get
$$\la\x^2(\alpha)\ra = 6\int dR\;\, C_2(R\n_1,\n_1)C_2(R\n_2,\n_2)
                                    C_2(R\n_3,\n_3) \;.\eqno(20)$$
When $\alpha=0$, equation~(20) reduces to equation~(11).

The value of $\la\x^2(\alpha)\ra{}^{1/2}$ can be computed by doing the
integral in equation~(20) numerically.  The result is plotted in Figure~1,
assuming an $n=1$ power spectrum, the COBE window function
($7^\circ\llap.5$~FWHM beam, $l=0,1,2$ terms removed),
and full sky coverage.  This indicates the magnitude of $\x(\alpha)$
which is to be expected in the COBE data due to cosmic variance, even though
the ensemble average of $\x(\alpha)$ is (assumed to be) zero.  The experimental
value of $\x(\alpha)$ would have to rise significantly above the curve in
Figure~1 in order to indicate a departure from gaussian statistics.  The
restriction to $|b|>20^\circ$ in the COBE data would enhance $\vx{}^{1/2}$
by a factor of 1.23 for small values of $\alpha$, and by a somewhat larger
factor (which is difficult to estimate reliably) for $\alpha\gtwid20^\circ$.
Note again that while the probability distribution for $\x(\alpha)$ must be
symmetric about $\x(\alpha)=0$, it will not be gaussian, and so
$\pm\vx{}^{1/2}$ does not represent a true 68\% confidence interval.

As an example of the utility of this result, let us examine the effect of
cosmic variance on the prediction of inflationary models for the three-point
correlation function.  For an $n=1$ power spectrum, the angular dependence of
the three-point function is determined, but the overall amplitude is model
dependent (Falk \etal\ 1993). In single-field inflation models, the amplitude
of the three-point function is proportional to $\mu/H$, where $\mu$ is the
coefficient of the cubic term in the potential of the inflaton field, and $H$
is the Hubble constant during inflation.  In two-field models (e.g., Kofman
\etal\ 1991), this amplitude is proportional to the strength of the coupling
between the two fields.  Figure~2 shows the ensemble average of the
three-point correlation function, assuming the COBE window function (Falk
\etal\ 1993).  The overall amplitude is that of a single-field model with
$\mu/H=0.1$.  The shaded area indicates a band of $\pm\vx{}^{1/2}$ about the
mean, assuming full sky coverage.  Of course, there will be corrections to
$\vx$ coming from the non-gaussian aspect of the distribution, but these will
be suppressed by a factor of $\mu^2/H^2$, and cannot significantly change the
result.  We see that unless $\mu/H$ is larger by at least a factor of two,
the wiggles at large $\alpha$ drop below the uncertainty due to cosmic
variance.  This is unfortunate; Luo \& Schramm (1993) have pointed out that
the height of the $\alpha=0$ peak depends sensitively on the power-spectrum
index $n$, but to determine $n$ from the three-point function requires
normalizing it at large $\alpha$.  This is impossible unless the three-point
function is large enough to raise its $\alpha\gtwid 20^\circ$ features well
above the uncertainty due to cosmic variance (to say nothing of the uncertainty
due to experimental noise, which is likely to be at least comparable).
Furthermore we see that we must have $\mu/H\gtwid 0.015$ in order to keep
even the $\alpha=0$ peak out of the cosmic noise.

To summarize, the fluctuations in the cosmic microwave background radiation
may contain deviations from gaussian statistics which would be reflected in a
nonzero value of three-point correlation function of $\Delta T$.  Any potential
observation of the three-point correlation function is limited by cosmic
variance, even if a whole-sky map of $\Delta T$ is available.  The cosmic
variance of the three-point correlation function depends on both the underlying
cosmological model and the experimental measurement strategy.  For COBE, the
skewness, defined as the sky average of $(\Delta T)^3$ with the dipole
and quadrupole terms removed from $\Delta T$, must exceed $1.3\,Q^3$
in order to be indicative of non-gaussian fluctuations.  For small-scale
experiments, the skewness must exceed $(4\pi/A)^{1/2}\theta_{\rm c}C_0^{3/2}$.

I thank Douglas Scott and Martin White for helpful comments.
This work was supported in part by NSF Grant Nos.~PHY--91--16964
and AST--91--20005.

\vfill\eject
\hoffset=0.25truein
\hsize=6.25truein
\parindent=-0.25truein

\centerline{REFERENCES}
\vskip0.1in

{\frenchspacing

Abbott, L. F., \& Wise, M. B. 1984a, Phys. Lett., B135, 279

Abbott, L. F., \& Wise, M. B. 1984b, ApJ, 282, L47

Bond, J. R., \& Efstathiou, G. 1987, MNRAS, 226, 655

Cayon, L., Martinez-Gonzalez, E., \& Sanz, J. L. 1991, MNRAS, 253, 599

Falk, T., Rangarajan, R., and Srednicki, M. 1993, ApJ, 403, L1

Kofman, L., Blumenthal, G.~R., Hodges, H.~M., \& Primack, J.~R. 1991,
in Large Scale Structures and Peculiar Motions in the Universe,
ed. D. W. Latham \& L. A. Nicolaci da Costa (ASP Conf. Ser., 15), 339

Luo, X., and Schramm, D.~N. 1993, Phys. Rev. Lett., submitted, astro-ph/9305009

Peebles, P. J. E. 1982, ApJ, 263, L1

Scaramella, R., \& Vittorio, N. 1990, ApJ, 353, 372

Scaramella, R., \& Vittorio, N. 1991, ApJ, 375, 439

Scott, D., Srednicki, M., and White, M. 1993, ApJ, submitted, astro-ph/9305030

Smoot, G.~F., \etal\ 1992, ApJ, 396, L1

White, M., Krauss, L., \& Silk, J. 1993, ApJ, submitted, astro-ph/9303009
}

\vskip0.2in
\vskip\parskip

\centerline{FIGURE CAPTIONS}
\vskip0.1in

Fig.~1.---The square root of the cosmic variance of the the three-point
correlation function $\vx{}^{1/2}$ as a function of the beam-separation
angle~$\alpha$, in units of $Q^3$ (where $Q$ is the r.m.s.~ensemble average
of the quadrupole moment), assuming an $n=1$ spectrum of fluctuations, the
COBE window function ($7^\circ\llap.5$~FWHM beam, $l=0,1,2$ terms removed),
and full sky coverage.

Fig.~2.---The prediction of inflation for the three-point correlation function
$\la\x(\alpha)\ra$ as a function of the beam-separation angle~$\alpha$,
in units of $Q^3$, assuming an $n=1$ spectrum of fluctuations and
the COBE window function.  The overall amplitude is that of a single-field
inflation model specified by $\mu/H=0.1$.  The gray band indicates a range
of $\pm\vx{}^{1/2}$ about the central value, assuming full sky coverage.

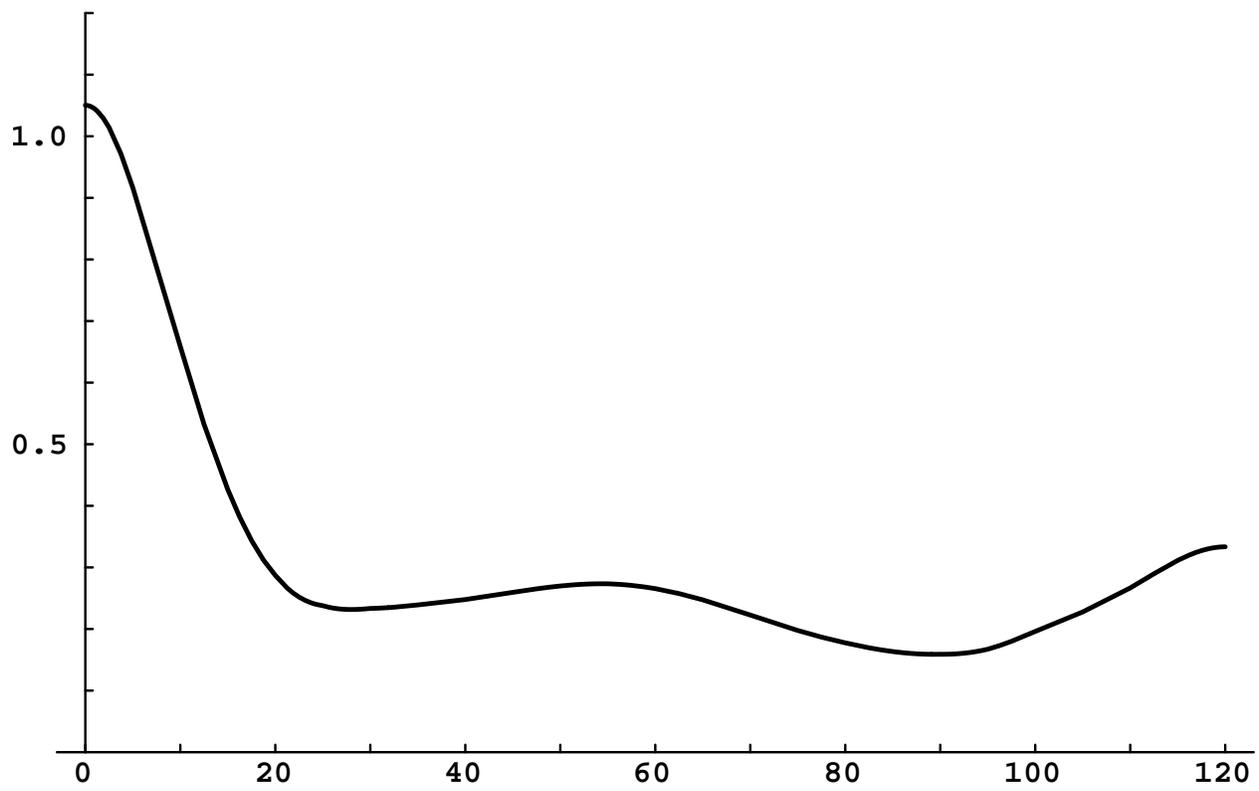
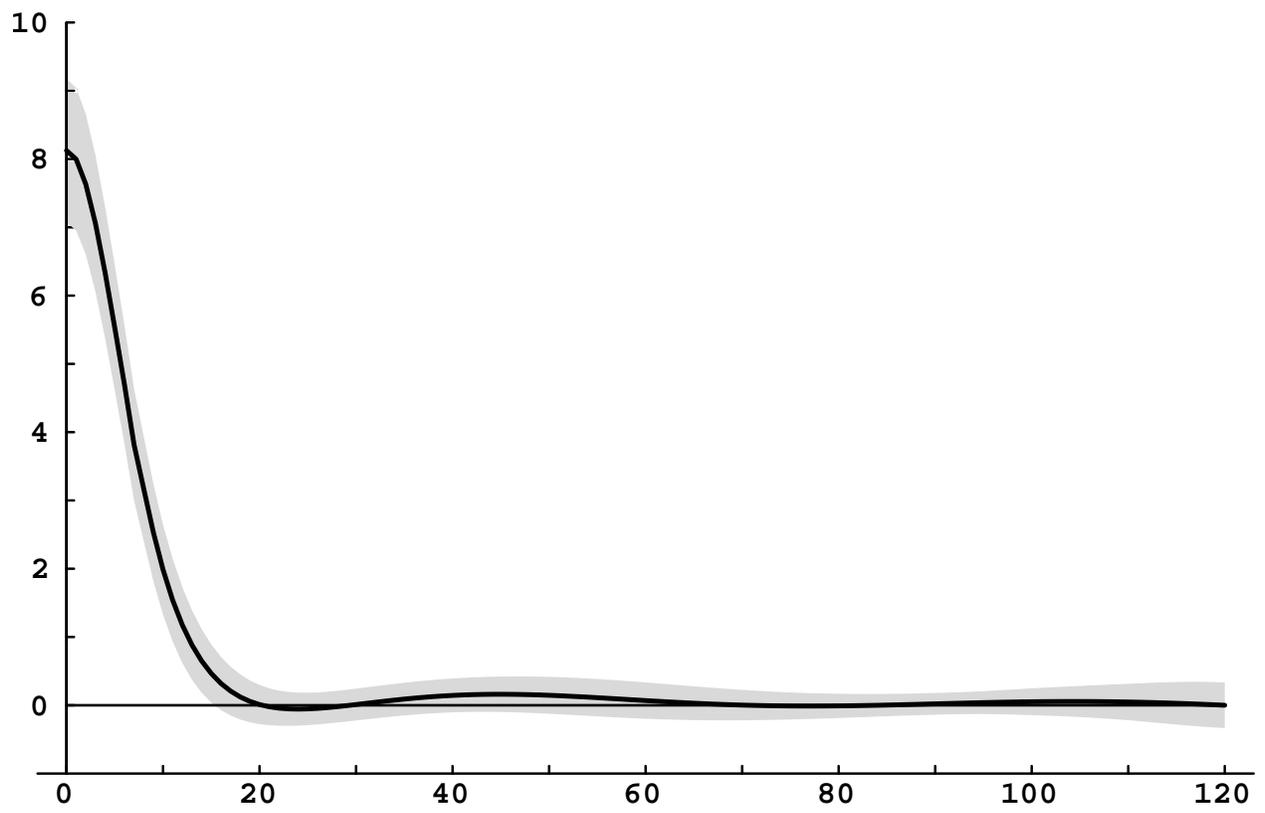
\end